\documentstyle[aps,prl,epsfig,multicol,mcolm]{revtex}
\begin{document}

\draft
\preprint{SNUTP-97-137, MSC-97-03}

\title{Quantum Phase Transitions in Josephson-Junction Chains}
\author{M.-S. Choi$^1$, J. Yi$^2$, M.Y. Choi$^2$, J. Choi$^3$, and
S.-I. Lee$^1$}
\address{$^1$ Department of Physics, Pohang University of Science and
Technology, Pohang 790-784, Korea}
\address{$^2$ Department of Physics and Center for Theoretical
Physics, Seoul National University, Seoul 151-742, Korea}
\address{$^3$ Department of Physics, Keimyung University, Taegu
704-701, Korea}
\date{Oct.~20 1997}
\maketitle

\begin{abstract}
We investigate the quantum phase transition 
in a one-dimensional chain of ultra-small superconducting
grains, considering both the self- and junction capacitances.  
At zero temperature, 
the system is transformed into a two-dimensional system of classical
vortices, where the junction capacitance introduces anisotropy in the
interaction between vortices.
This leads to the superconductor-insulator transition of the
Berezinskii-Kosterlitz-Thouless type, as the ratios of
the Josephson coupling energy to the charging energies are varied. 
It is found that the junction capacitance plays a role similar
to that of dissipation and tends to suppress quantum
fluctuations; nevertheless the insulator region survives even for
arbitrarily large values of the junction capacitance.

\end{abstract}

\newcommand {\varN}	{ {\cal N} }
\newcommand {\varO}	{ {\cal O} }
\newcommand {\half} { \frac{1}{2} }


\begin{multicols}{2}

Quantum phase transitions, which are induced by quantum fluctuations at
zero temperature, are distinguished from classical phase
transitions in several important
respects; this has attracted much attention in recent years~\cite{Sondhi97}.
In particular,
advances in fabrication techniques have made available
arrays of ultra-small superconducting grains,
where the charging energy dominates the
Josephson coupling energy and accordingly, quantum fluctuation effects
are of paramount importance.  
Such Josephson-junction arrays
have become a prototype system displaying quantum phase transitions
between the superconducting and the insulating phases.
In the vicinity of the superconductor-insulator transition,
the fluctuation effects depend crucially
on the dimensionality of the system.
In the case of two-dimensional (2D) arrays,
rich effects of quantum fluctuations and resulting phase transitions 
have been examined for a rather general form of the capacitance matrix,
although there still exist unsettled issues
in the quantum regime, such as low-temperature
re-entrance~\cite{Faziox91,BJKim95,Rojas}.
On the other hand, one-dimensional (1D) chains of Josephson junctions, 
where quantum fluctuations should be more important, have been
studied mainly in the two limiting cases: the self-charging model and 
the nearest-neighbor model where only nearest neighboring 
charges interact~\cite{Bradle84,end_note:1}. 
In the 1D system with only self-capacitance, the role of
dissipation on the quantum phase transition~\cite{Bobber92} 
as well as the persistence current and voltage~\cite{Choi93} 
has also been considered.

This paper investigates the quantum phase transitions 
in general Josephson-junction chains with
{\em both the self- and junction capacitances}. 
At zero temperature, 
the system is transformed into a two-dimensional system of classical
vortices, where the junction capacitance introduces anisotropy in the
interaction between vortices.
This leads to the superconductor-insulator transition of the
Berezinskii-Kosterlitz-Thouless (BKT) type~\cite{Berezi71,Koster74}, 
as the ratios of
the Josephson coupling energy to the charging energies are varied. 
Interestingly, the junction capacitance here plays a role similar
to that of dissipation and tends to suppress quantum
fluctuations, enhancing superconductivity.  
However, the suppression is not strong enough, and
the insulator region still remains even for
arbitrarily large values of the junction capacitance.

We consider a one-dimensional array of $N$ ultra-small superconducting grains, 
each of which is coupled to nearest-neighboring grains via Josephson junctions.
The system is characterized by three energy scales,
Josephson coupling energy $E_J$ and charging
energies $E_{0}\equiv{}e^2/2C_0$ and $E_{1}\equiv{}e^2/2C_1$,
where $C_0$ is the self-capacitance of each grain and 
$C_1$ is the junction capacitance between nearest-neighboring grains.
It is described by the Hamiltonian
\begin{equation}
H
= \frac{1}{2K_0}\sum_{i,j=1}^N n_i C^{-1}_{ij} n_j
  - K_0\sum_{i=1}^N\cos(\phi_{i+1}-\phi_i)
  ,
  \label{QPM}
\end{equation}
where the number $n_i$ of Cooper pairs and
the phase $\phi_i$ of the superconducting order parameter at site $i$
are quantum-mechanically conjugate variables: $[n_i,\phi_j]=i\delta_{ij}$,
and the energy has been
rescaled in units of the Josephson plasma frequency
$\hbar\omega_p\equiv\sqrt{8E_{0}E_J}$.
Here we have introduced $K_0\equiv\sqrt{E_J/8E_{0}}$ to describe the case 
$C_0 \neq 0$, and written
the capacitance matrix in the form
$$
C_{ij}
= \delta_{ij} + \lambda^2\left(
  2\delta_{i,j}-\delta_{i,j-1}-\delta_{i,j+1} \right)
  , \label{Cij:1}
$$
where $\lambda^2\equiv{}C_1/C_0$.
On the other hand, when $C_0 =0$, 
it is convenient to introduce $K_1\equiv\sqrt{E_J/8E_{1}}$.
Note that $\lambda$ can also be written in the form: $\lambda =K_1/K_0$.

In the imaginary-time path-integral representation, the partition
function of the system reads
\begin{equation}
Z
= \prod_{j,\tau}\sum_{n_{j,\tau}}
  \int_0^{2\pi}\frac{d\phi_{j,\tau}}{2\pi}\;
  \exp\left\{-\sum_{\tau=0}^{\beta-1}L[n,\phi]\right\}
\end{equation}
with the action
\onecolm
\begin{equation}
L[n,\phi]
= i\sum_{j}^N n_{j,\tau}\partial_\tau\phi_{j,\tau}
  + \frac{1}{2K_0}\sum_{i,j}^N n_{i,\tau}C^{-1}_{ij} n_{j,\tau}
  - K_0\sum_{j}^N \cos\partial_x\phi_{j,\tau}
  ,
  \label{QPM:L}
\end{equation}
\twocolm\noindent
where the temperature has been rescaled according to
$\hbar\omega_p\beta\to\beta$.   
In Eq.~(\ref{QPM:L}), $\partial_x$ and $\partial_\tau$ denote 
the difference operator with respect to
the position $x$ and to the imaginary time $\tau$, respectively:
$\partial_x\phi_{j,\tau}\equiv \phi_{j+1,\tau}-\phi_{j,\tau}$ and
$\partial_\tau\phi_{j,\tau}\equiv \phi_{j,\tau+1}-\phi_{j,\tau}$,
and the (imaginary) time slice $\delta\tau$ in the
interval $[0,\beta]$ has been chosen to be unity (in units of
$\hbar\omega_p$)~\cite{end_note:2}.
Our main concern here is the quantum phase transition at zero temperature,
and accordingly, the limit $\beta\to\infty$ as well as the thermodynamic
limit $N\to\infty$ is to be taken.
We then apply the Villain approximation~\cite{Josexx77} and
integrate out $\{\phi_{j,\tau}\}$, to obtain
the partition function in terms of an additional set of integer variables
$\{m_{j,\tau}\}$ (as well as $\{n_{j,\tau}\}$).
The corresponding action is given by
\begin{equation}
L[n,m]
=  \frac{1}{2K_0}\sum_{i,j}n_{i,\tau}C^{-1}_{ij} n_{j,\tau}
  + \frac{1}{2K_0}\sum_jm_{j,\tau}^2
  ,
  \label{L:1}
\end{equation}
where the two sets of integer variables satisfy the constraint
$\partial_x m_{j,\tau}+\partial_\tau{}n_{j,\tau}=0$.
This constraint is conveniently taken into account by introducing an integer
field $A_{\tilde{j},\tilde\tau}$ defined on the space-time dual
lattice $(\tilde{j},\tilde\tau)\equiv(j+1/2,\tau+1/2)$ in such a way that
$m_{j,\tau}=-\partial_\tau{}A_{\tilde{j},\tilde\tau}$ and 
$n_{j,\tau}=\partial_x A_{\tilde{j},\tilde\tau}$.
Henceforth we will work on the dual lattice, and drop for simplicity
the tilde sign over site indices.
The partition function is thus written in terms of the unconstrained
summation of $\exp\{-\sum_\tau L[A]\}$ over $A_{j,\tau}$'s,
with the action
\onecolm
\begin{equation}
 L[A]
=  \frac{1}{2K_0}\sum_{i,j}(\partial_xA_{j,\tau})
    C^{-1}_{ij}(\partial_xA_{j,\tau})
  + \frac{1}{2K_0}\sum_j(\partial_\tau A_{j,\tau})^2
  .
\end{equation}
\twocolm

Finally, the Poisson summation formula, which decomposes
$A_{j,\tau}$ into a real-valued field and a new integer
field $v_{j,\tau}$, allows us to integrate out the real-valued field 
$A_{j,\tau}$.
This leads, apart from the spin-wave part, 
to the 2D system of classical vortices described by the Hamiltonian
\begin{equation}
H_{v}
= 2\pi^2 K_0 \sum_{i\tau,j\sigma}
  v_{i,\tau}\,U(i{-}j,\tau{-}\sigma)\,v_{j,\sigma},
  \label{2DVG}
\end{equation}
where the vortex interaction is given by
\begin{eqnarray}
U(x,\tau)
& = & \int_0^{2\pi}\frac{dq}{2\pi}
  \int_0^{2\pi}\frac{d\omega}{2\pi}\;
  e^{+iqx-i\omega\tau}\:\widetilde{U}(q,\omega) \nonumber
  ,
  \\
\widetilde{U}(q,\omega)
& = & \left[\frac{\Delta(q)}{\widetilde{C}(q)} + \Delta(\omega)\right]^{-1}
  \label{U}
\end{eqnarray}
with the Fourier transforms of the lattice Laplacian 
$\Delta(z)= 2(1-\cos{z})$ and 
of the capacitance $\widetilde{C}(q)=1+\lambda^2\Delta(q)$.
Note that unless $C_0=0$, the diagonal piece
$U_0\equiv U(x{=}0,\tau{=}0)$ becomes arbitrarily large 
as $N$ and $\beta$ are increased to infinity.  
Therefore the diagonal term $2\pi^2 K_0 U_0 (\sum_{j,\tau} v_{j,\tau})^2$
should vanish in Eq.~(\ref{2DVG}), leaving the vorticity neutrality condition:
$\sum_{j,\tau} v_{j,\tau} =0$,
and the Hamiltonian (\ref{2DVG}) may be written in the form
\begin{equation}
H_{v}
= -\pi K_0 \sum_{i\tau,j\sigma}
  v_{i,\tau}\,\widehat{U}(i{-}j,\tau{-}\sigma)\,v_{j,\sigma}
  \label{2DVG1}
\end{equation}
with $\widehat{U}(x,\tau)\equiv{}2\pi[U_0 -U(x,\tau)]$.
The behavior of the (reduced) vortex interaction $\widehat{U}(x,\tau)$ 
for several values of $\lambda$ 
is displayed in Figs.~\ref{fig:U:x,tau} and \ref{fig:U:contour}, 
which manifests the logarithmic behavior at large length scales.
In particular, the short-range anisotropy prominent for large values
of $\lambda$ decreases rapidly with the distance.
According to the renormalization group (RG)
theory of the 2D Coulomb gas, such short-range anisotropy should not
affect the universality class.  Thus in the
spirit of the RG theory, the system is expected to exhibit qualitatively 
the same critical behavior as the 2D Coulomb gas~\cite{Minnha87}.

We now investigate the quantum phase transitions displayed by
the Hamiltonian in Eq.~(\ref{2DVG}).  
First we consider the simplest case of $C_1=0$, which has been studied
in Ref.~\cite{Bradle84}.  In this self-charging limit, the interaction in
Eq.~(\ref{U}) takes the simple form
$\widetilde{U}(q,\omega)=[\Delta(q)+\Delta(\omega)]^{-1}$ or
$\widehat{U}(x,\tau)\approx \ln\sqrt{x^2+\tau^2}+\frac{3}{2}\ln{2}+\gamma_E$,
where $\gamma_E$ is the Euler number.  Thus the Hamiltonian (\ref{2DVG1})
describes precisely the isotropic 2D Coulomb gas, and the system
undergoes a BKT transition~\cite{Berezi71,Koster74,Josexx77}
from the insulating phase to the superconducting one as $K_0$ is
increased~\cite{Bradle84}.

We next consider the opposite limit $C_0=0$, where
the relevant (dimensionless) coupling constant is $K_1$ instead of $K_0$.
Accordingly, it is proper to
use $\hbar\omega_p\equiv\sqrt{8E_{1}E_J}$ in rescaling
the energy and the imaginary time,
which in turn gives
$\widetilde{C}(q)=\lambda^{-2}+\Delta(q)$ and
$\widetilde{U}(q,\omega)=[1 + \Delta(\omega)]^{-1}$.
As a result, we obtain the Hamiltonian in the form
\begin{equation}
H_{v}
= 2\pi^2 K_1 \sum_{i\tau,j\sigma}
  v_{i,\tau} U(i{-}j,\tau{-}\sigma)v_{j,\sigma}
  ,
  \label{2DVG:2}
\end{equation}
where, in sharp contrast to the previous self-charging case, 
the vortex interaction $\widehat{U}(x,\tau)\equiv{}2\pi[U_0 -U(x,\tau)]$ 
is short-ranged:
\begin{equation}
\widehat{U}(x,\tau)
\approx \frac{2\pi}{\sqrt{5}}\left[
    1-\delta_{x,0}e^{-|\tau|}
  \right]
\end{equation}
for $|\tau|\gg 1$.
It is further of particular importance 
that $U_0$ in general does not diverge,
which implies that the diagonal term in the Hamiltonian~(\ref{2DVG:2}) 
does not give the vorticity neutrality condition.
Thus (unbound) free vortices become pervasive,
and the system remains insulating for any nonzero value of $K_1$.
This can also be understood in the charge
representation, where the partition function reads
\begin{equation}
Z
= \prod_{j,\tau}\sum_{n_{j,\tau}}\exp\left\{
    -\frac{1}{\beta N}\sum_{q,\omega}
    |n(q,\omega)|^2G^{-1}(q,\omega)
  \right\}
  ,
  \label{charge-rep}
\end{equation}
with $G(q,\omega) = \Delta(q)[\Delta(\omega)+1]^{-1}$.
It follows from the analytic continuation of $G(q,\omega)$
that the charge excitation has a gap $E_g\sim\hbar\omega_p$.
Thus the long-range interaction $C_{ij}^{-1}$ between
charges gives rise to the gap, resulting in an insulator.
It is also of interest to compare this case with the
``nearest-neighbor model'' in Ref.~\cite{Bradle84}: In the latter,
only the nearest-neighboring charges interact, 
which allows only bound vortices.  Consequently, 
the system is superconducting
for all finite values of the interaction strength.

We now turn to the more realistic case that $\lambda$ is small but nonzero,
i.e., $C_0\gg{}C_1$.
In this case, the vortex interaction
$\widehat{U}$ is, to the order of $\lambda^2$, isotropic
and has the asymptotic behavior:
$\widehat{U}(x,\tau)
 \approx \ln\sqrt{x^2 +\tau^2} + \varepsilon(\lambda)$,
where $\varepsilon(\lambda)$ is the vortex pair creation energy (per vortex) 
given by
$\varepsilon(\lambda) 
 \approx (3/2)\ln{2} + \gamma_E + (\pi{-}1)\lambda^2 + \varO(\lambda^4)
$.
Hence the system again reduces to the 2D Coulomb gas, with
the vortex fugacity
$y(\lambda)\equiv{}\exp[-2\pi{K_0}\varepsilon(\lambda)]$ diminished by
the factor $\exp[-2\pi{K_0}(\pi-1)\lambda^2]$.
The standard RG theory of the BKT
transition~\cite{Koster74,Josexx77} then predicts the transition point
$K_0^c(\lambda)$ slightly decreased from $K_0^c(0)$ by the amount
\begin{equation}
\frac{\delta K_0^c}{K_0^c(0)}
 = -(\pi-1)\, \frac{2\pi K_0^c(0)-4}{2\pi K_0^c(0)-3}\, \lambda^2
  + \varO(\lambda^4)
 \approx -1.6 \lambda^2
  .
\end{equation}

In this limit ($\lambda\ll{}1$), the Hamiltonian in Eq.~(\ref{QPM}) can
also be represented in terms of phase variables, which yields
a 2D XY model with an additional interaction for nonzero $\lambda$.
To see this, we neglect ${\cal O}(\lambda^4)$, and 
write the inverse capacitance matrix in the form
\begin{equation}
C_{ij}^{-1}\approx (1-2\lambda^{2})\delta_{ij}
        +\lambda^{2}(\delta_{i,j+1}+\delta_{i,j-1}).
\end{equation}
Note that without the off-diagonal term in $C_{ij}^{-1}$, 
the first and the second terms in the action (\ref{QPM:L}) 
would just be the Villain form of the cosine action along the
$\tau$ direction~\cite{Bradle84,Choi93}.
To examine the effects of the off-diagonal term, we use the identity
\onecolm
$$
\exp\left[\frac{\lambda^2}{4K_0}(\partial_x n_{j,\tau})^{2}\right]
 =\frac{1}{\sqrt{\pi}}\int_{-\infty}^{\infty}dz_{j,\tau}
\exp\left[-z_{j,\tau}^{2}
 -\frac{\lambda}{\sqrt{K_0}}\,z_{j,\tau}\, \partial_{x}n_{j,\tau}\right],
$$
\twocolm\noindent
which allows to separate the charge variables at different sites.
The resulting action then takes the Villain form 
of the cosine action with the argument
$\partial_\tau\phi_{j,\tau}{+} i\sqrt{\lambda^2/K_0}\, \partial_x z_{j,\tau}$ 
along the $\tau$ direction.
To the order of $\lambda^2$,
the Gaussian integration over $z_{j,\tau}$ 
leads to the effective phase Hamiltonian
\onecolm
\begin{eqnarray}
H_{\rm eff}[\phi]
& = & -K_0 \left(\frac{1}{\sqrt{1-\lambda^2}}+\frac{\lambda^2}{4K_0}\right)
    \sum_{j,\tau}[\cos(\partial_x\phi_{j,\tau})
    +\cos(\partial_\tau\phi_{j,\tau})
    ]
  \nonumber \\
& & \mbox{}
  +\lambda^2 K_0
  \sum_{j,\tau}[\partial_\tau\sin\partial_x\phi_{j,\tau}]^2
  ,
  \label{XYOhmic}
\end{eqnarray}
\twocolm\noindent
where the anisotropy has been removed by rescaling again the $\tau$ axis
by the factor $(1{-}\lambda^2)^{-1/2}{+}\lambda^2/4K_0$.
Note that Eq.~(\ref{XYOhmic}) reduces, for $\lambda=0$, 
to the standard 2D XY Hamiltonian.
Here the junction capacitance not only enhances the effective coupling
of the XY model but also introduces an additional interaction given by
the second term in Eq.~(\ref{XYOhmic}).
Interestingly, the latter is very similar in form to the 
dissipation term in the effective action, which is known 
to suppress quantum fluctuations~\cite{BJKim95,Bobber92}.
Therefore both effects contribute to the enhancement of phase coherence,
and it is concluded that
the junction capacitance in general tends to
suppress quantum fluctuations induced by the self-capacitance, 
thus reducing the insulator region.

For larger values of $\lambda$, 
Figs.~\ref{fig:U:x,tau} and \ref{fig:U:contour}(b) show that the vortex
creation energy $\varepsilon(\lambda)$ and the 
short-range behavior of the interaction $\widehat{U}(x,\tau)$ 
are highly anisotropic.
Still at large length/time scales,
$\widehat{U}(x,\tau)$ becomes isotropic and logarithmic, which can also
be confirmed by the asymptotic expansion for large $\tau$:
\onecolm
\begin{equation}
\widehat{U}(0,\tau)
= \frac{2\pi}{\sqrt{5}}\lambda
  + \ln\tau 
  + \left\{\gamma_E + 2\ln{2}-\half\ln{5}-\frac{127}{150}\right\}
  + \varO(1/\tau,1/\lambda^{2})
  .
  \label{eq:4}
\end{equation}
\twocolm\noindent
Although the anisotropy in the short-range
behavior of $\widehat{U}$ may slightly alter the details of the RG flow,
the qualitative features at large length scales are expected unaffected. 
Thus it is concluded that for any value of
finite $\lambda$ the system undergoes the superconductor-insulator
transition of the BKT universality class.

Figure~\ref{fig:ph_diag} displays the schematic phase diagram 
on the $K_0$-$K_1$ plane.
Although the precise phase boundary, i.e., 
the detailed behavior of $K_0^c(\lambda)$,
in general depends on the microscopic length scales such as
the (imaginary) time slice $\delta\tau$,
the universal (large-length scale) behavior should not be affected.
Thus the phase boundary
between the superconducting and the insulating phases 
is concluded to belong to the BKT universality class.
As $\lambda \,(\equiv K_1/K_0)$ is increased,
the vortex creation energy also grows monotonically,
resulting in the decrease of $K_0^c(\lambda)$
toward the limiting value $2/\pi$.
(Note that the value $2/\pi$ corresponds to the zero vortex fugacity
or the infinite vortex creation energy.)
This lowering of the transition point and the resulting enhancement
of superconductivity reflects the role of
the junction capacitance: It tends to suppress the quantum fluctuations
induced by the self-capacitance,
as manifested in the phase representation given by Eq.~(\ref{XYOhmic}).
There has been few experimental studies of quantum phase transitions
in 1D Josephson jucntion arrays.  Recent interest in 1D tunnel
junction arrays~\cite{Delsin96} and advances in submicron fabrication
techniques strongly motivate experimenters in this field.

MSC is grateful to W.~G. Choe, G.~S. Jeon, and K.-H.
Wagenblast for helpful discussions.  This work was supported in part by the
Basic Science Research Institute Program, Ministry of Education of
Korea and in part by the Korea Science and Engineering Foundation through the
SRC Program. 


\begin{thebibliography}{10}
\bibitem{end_note:1}
It should be noted that the second case (the nearest-neighbor model in
Ref.~\cite{Bradle84}) is rather unrealistic and in particular does not
correspond to the case $C_0=0$ and $C_1\neq 0$ shown schematically in
Fig.~2 of Ref.~\cite{Bradle84}. In the latter case, the charge
interaction $C^{-1}_{ij}$ is long-ranged rather than between nearest
neighbors.
\bibitem{end_note:2}
The critical behavior of the system should not be affected by the
choice of $\delta\tau$ (see Ref.~\cite{Sondhi97}). Here all the
dynamics of the system occurs over the time scale $1/\omega_p$, making
it proper to choose $\delta\tau\sim\hbar\omega_p$.
\end{thebibliography}

\narrowtext  

\begin{figure}
\begin{center}
\epsfig{file=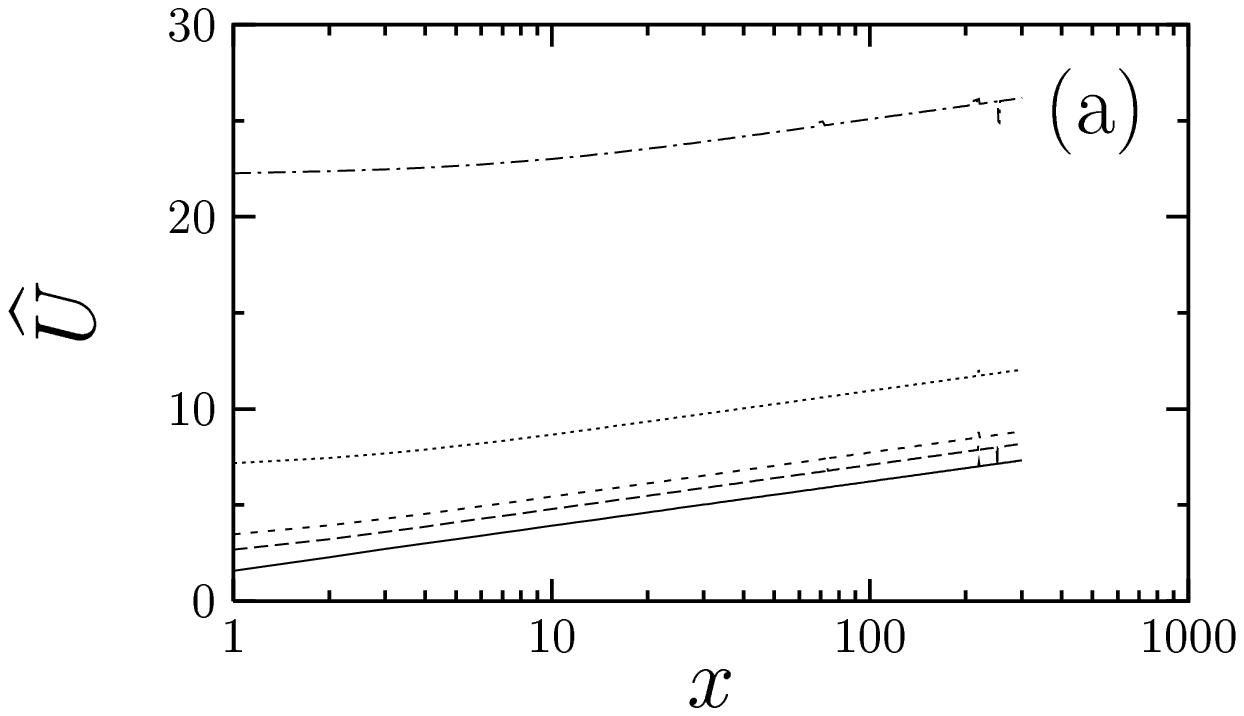,clip=,width=80mm} \\
\epsfig{file=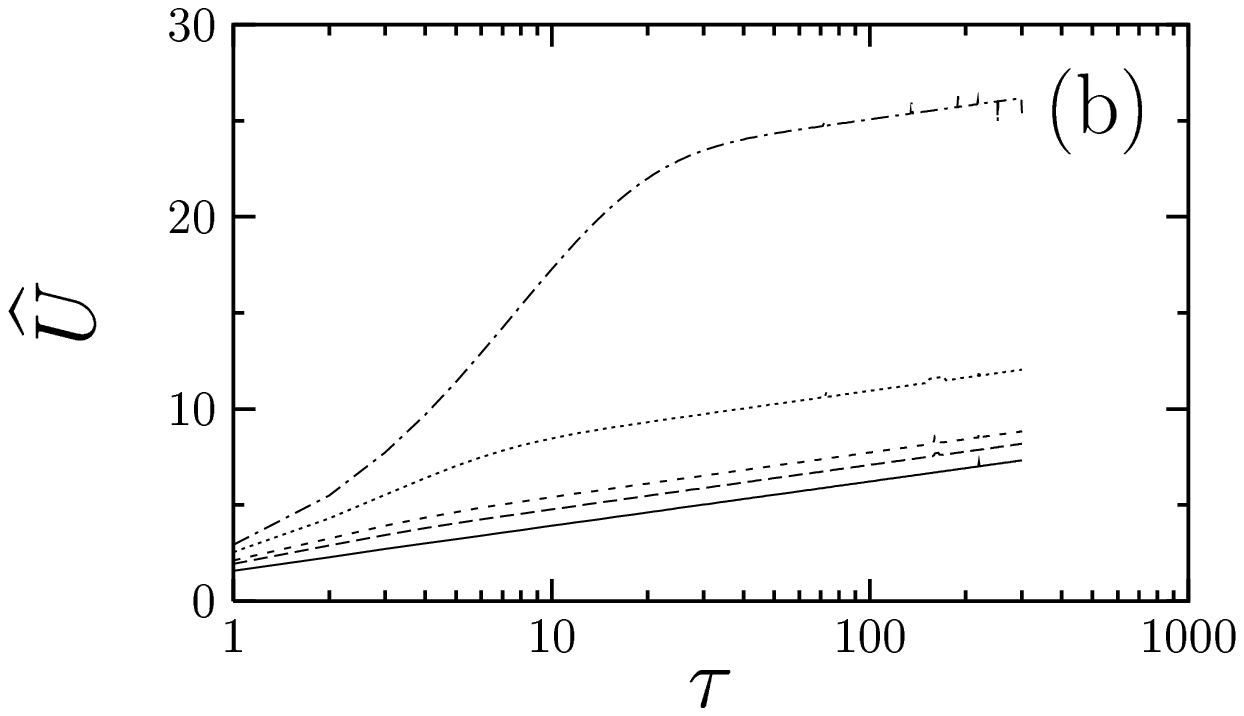,clip=,width=80mm}
\end{center}
\caption{Vortex interaction potential $\widehat{U}$ along
(a) the $x$-axis and (b) the $\tau$-axis for 
$\lambda=0,\, 0.5,\, 1,\, 5$, and $50$ (from the bottom to the top).}
\label{fig:U:x,tau}
\end{figure}

\begin{figure}
\begin{center}
\epsfig{file=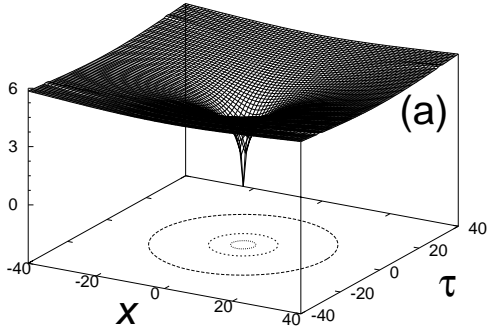,clip=,width=80mm} \\  
\epsfig{file=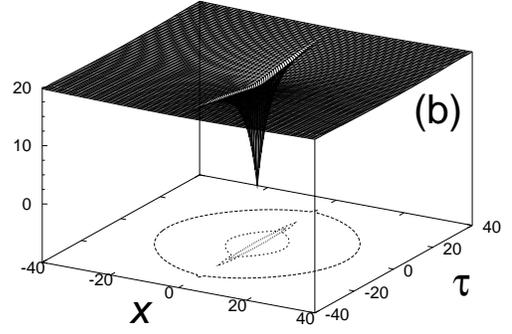,clip=,width=80mm} 
\end{center}
\caption{Plot of the vortex interaction $\widehat{U}(x,\tau)$ 
for (a) $\lambda^2=0.1$ and (b) $\lambda^2=30$.}
\label{fig:U:contour}
\end{figure}

\begin{figure}
\centerline{\epsfig{file=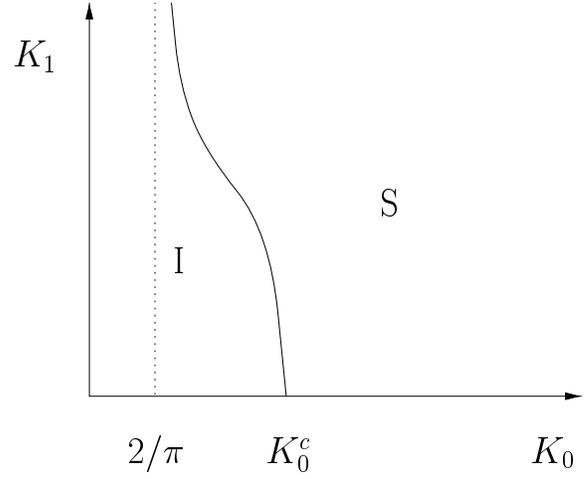,clip=,width=80mm}}
\caption{Schematic phase diagram of a Josephson-junction chain at
zero temperature.  As $\lambda$ gets increased, $K_0^c$ decreases toward
the limiting value $2/\pi$.}
\label{fig:ph_diag}
\end{figure}

\end{multicols}

\end{document}